\documentclass[12pt]{article}
\usepackage{graphicx}
\def\bfT2{\mathord{\mbox{\boldmath $T$}^2}}

\begin{document}
\begin{center}
{\bf STICKINESS OF TRAJECTORIES IN A PERTURBED ANOSOV SYSTEM}
\end{center}

\begin{center}
{\bf G.M. Zaslavsky$^{(1,2)}$  and M. Edelman$^{(1)}$ \\
$^{(1)}$Courant Institute of Mathematical Sciences,
New York University,
251 Mercer St., New York, NY 10012, USA \\
$^{(2)}$Department of Physics, New York University,
2-4 Washington Place, New York, NY 10003, USA\\
Email: zaslav@cims.nyu.edu\\
Email: edelman@cims.nyu.edu
}
\end{center}

\begin{abstract}
We consider a perturbation of the Anosov-type system, which leads to the
appearance of a hierarchical set of islands-around-islands. We
demonstrate by simulation that the boundaries of the islands are sticky to
trajectories. This phenomenon leads to the distribution of
Poincar\'{e} recurrences
with power-like tails in contrast to the exponential distribution in the
Anosov-type systems.
\end{abstract}
\newpage
\noindent
{\bf 1. INTRODUCTION}

Anosov maps (or Anosov-type maps, or Anosov diffeomorphisms)
can be considered as a strong
idealization of realistic systems.
The reason of this is that  Hamiltonian systems are
not uniformly hyperbolic and
their phase space features a complex mixture of zones of different
dynamics. The boundaries of the zones influence the trajectories and
drastically change asymptotic properties of
transport.  The ways how the Anosov
systems can lose their hyperbolicity is the subject of various
investigations \cite{Le,Pr,En,Ler,DC}. 
The main questions discussed here are a description
of the topological pattern of the destroyed hyperbolicity and the
description of the consequences of that loss. For the latter issue, it
could be, for example, the changes in particles transport
due to the changes
in the phase space topology
that occurs in Hamiltonian systems with mixed phase
space \cite{Z2002}, i.e. phase space with chaotic sea and islands.

In this paper we make few comments related to the
destruction of hyperbolicity and to the problem of transport. The considered
system is similar to the introduced one in \cite{Le}. It is a map 
that acts on the torus
${\bf T}^2$ and it depends on an intrinsic parameter $K \in (-\infty ,
\infty )$, and on the perturbation parameter $\epsilon_0$. The uniform
hyperbolicity exists when $\epsilon_0  = 0$ and in a region ${\bf K}_+$ of
the values of $K$. The system is most sensitive to the perturbation in the
vicinity of edges of ${\bf K}_+$. In the extended space
${\bf T}^2 \times {\bf K}$ the system is unstable with respect to an
arbitrary small $\epsilon_0$. Increasing $\epsilon_0$ leads to a sequence
of bifurcations. We demonstrate by simulations the appearance of a
hierarchical set of islands in the stochastic sea and a stickiness of their
borders. In other words, trajectories spend enormously long time in the
vicinity of the islands' border making the transport to be anomalous.
The ``anomalous'' means that the wandering of trajectories
strongly differs from the
random walk in the case of uniform hyperbolicity. This property is
established numerically by applying the method of $\epsilon$-separation
along reference trajectories developed in \cite{AZ,LZ,ZE2005}. 
The main result can be
expressed in a simplified form as follows: After the loss of hyperbolicity,
the considered Anosov system reveals the power law of the distribution of
Poincar\'{e} recurrences when $t$ (time) $\rightarrow \infty$, in contrast
to the exponential distribution for unperturbed Anosov system \cite{Ma}.

\noindent
{\bf 2. DESCRIPTION OF THE SYSTEM}

Consider a dynamical system $(x,y) \in {\bf T}^2$ defined
on the two-dimensional torus by the equation of motion
\begin{equation}
y_{n+1} = y_n + f(x_n ; \epsilon_0 ,K)  ; \ \ \ \
x_{n+1} = x_n + y_{n+1} \ , \ \ \ \ ({\rm modd} \ 1) 
\label{1}
\end{equation}
and a differentiable function
\begin{equation}
f(x_n ;\epsilon_0 , K ) = Kx_n - \epsilon_0 \psi (x_n ) 
\label{2}
\end{equation}
with two parameters $K$ and $\epsilon_0$. The map (\ref{1}) is
area-preserving for any
$f(x ;\epsilon_0 , K )$.
While the map acts in the phase space which is 
$\bfT2$, it doesn't transform the
boundary  of the torus $\partial \bfT2$ into  $\partial \bfT2$. Particularly, 
for $\epsilon_0 = 0$ it is evident since $K$ is not integer. This is a typical
physical situation. One can interpret the map  (\ref{1}) as a generator of 
trajectories $\{x_0, y_0; x_1, y_1; ...\}$ in the lifted space  $(x, y) 
\in (-\infty, \infty; -\infty, \infty )$, which should be, afterward, bended 
and put inside the compact phase space $(x, y)$ $modd 1$.
This way of the consideration of the map   (\ref{1}) permits to study infinite
diffusion in the lifted space and local measures, 
Poincar\'{e} recurrences measure, etc. in the phase space $T^2=(x,y)$ $modd 1$.

The eigenvalues of the tangent to (\ref{1}) map are
\begin{equation}
\lambda_{1,2} = \left( 1 + {1\over 2} K - {1\over 2} \epsilon_0
\psi^{\prime} (x) \right) \pm
\left[ \left( 1 + {1\over 2} K - {1\over 2} \epsilon_0 \psi^{\prime} (x)
\right)^2 - 1 \right]^{1/2} \ . 
\label{3}
\end{equation}
For $\epsilon_0 = 0$ the region
\begin{equation}
{\bf K}_+ : \ K < -4 \ , \ \ \ K > 0 \ ; \ \ \ \
(\epsilon_0 = 0) \label{4}
\end{equation}
corresponds to uniformly hyperbolic dynamics, and the region
\begin{equation}
{\bf K}_- : \ -4 < K < 0 \label{5}
\end{equation}
corresponds to periodic dynamics with zero Lyapunov exponent. For our
following consideration a specific form of $\psi (x)$ is not so important.
Nevertheless, we specify it for a convenience
\begin{equation}
\psi (x) = \sin x \cdot \cos^2 x \ , \label{6}
\end{equation}
as it was considered in \cite{Le,Ler}.

The system  (\ref{1}),  (\ref{2}) represents Anosov diffeomorphism for $\epsilon_0 = 0$,
$K \in {\bf K}_+$,
and it is structurally (topologically) stable for $\epsilon_0 \rightarrow 0$.
Our goal is to consider structural stability in the extended space
${\bf T}^2 \times {\bf K}$ where $K \in {\bf K}$.
\newpage
\noindent
{\bf 3. ISLANDS AND ISLANDS HIERARCHY}

\vspace{.2in}

\begin{figure}
\centering
\includegraphics[width=15 cm]{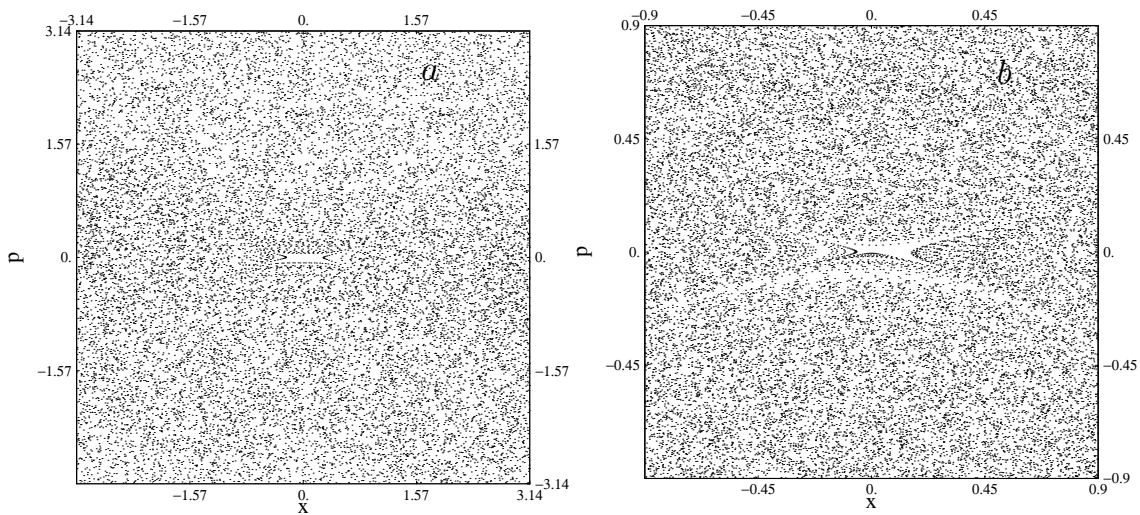}
\caption{\label{fig1} Poincar\'{e} map of a trajectory for the system (1) with (2), (6) and
$K = 0.2$, $\epsilon_0 = 0.2$ (left), $\epsilon_0 = 0.16$ (right). }
\end{figure}

{\it Property 1.} For arbitrary small $\epsilon_0$ there exists
$K^{\pm}_{\epsilon_0} \in {\bf K}$ such that for
$K \in (K^-_{\epsilon_0} , K^+_{\epsilon_0})$ phase space ${\bf T}^2$
has islands of periodic orbits that appear as a result of
the bifurcation.
The main reason for this is that the hyperbolicity has edges defined in
(\ref{4}),  (\ref{5} and arbitrary small $\epsilon_0$ influence the
system's stability
fairly close to the edges. An example of the island born is in Fig.~\ref{fig1}.
For $\epsilon_0 \neq 0$ the domain of stability
\begin{equation}
{\bf K}_- : \ -4 + \epsilon_0 \psi^{\prime} (x) < K < \epsilon_0
\psi^{\prime} (x) \label{7}
\end{equation}
depends on $x$, and the phase space becomes non-uniform.

For small $\epsilon_0$ the most interesting domain is near $x = 0$ and
small $K$. For small $x$ we have from (\ref{2}):
\begin{equation}
f(x) = \delta x + (7/6) \epsilon_0 x^3 + O(x^3 ) \ ; \ \ \
\delta = K - \epsilon_0 \ . \label{8}
\end{equation}
The map ( (\ref{1}), in the vicinity of (0,0), gives for $\delta\rightarrow 0$
$$
y_{n+2} = (1+\delta )y_n + 2\delta x_n + {7\over 6} \epsilon_0 x_n^3
+ O (\delta^2 )
$$
\begin{equation}
x_{n+2} = (1+3\delta )x_n + 2y_n + {7\over 3} \epsilon_0 x_n^3
+ O (\delta^2 )
\label{9}
\end{equation}
that is area-preserving up to the terms of the order $\delta^{3/2}$,
 $\delta x$. On the same level of accuracy, the map (\ref{9})
can be written as Hamiltonian equations
$$
\dot{y} = - {\partial H\over \partial x} \ , \ \ \ \
\dot{x} = - {\partial H\over \partial y} \ ,
$$
\begin{equation}
H \approx - {1\over 2} \delta x^2 + {1\over 2} y^2
+ O (|\delta |^{3/2} )
\label{10}
\end{equation}
$$
\dot{y} \approx {1\over 2} (y_{n+2} - y_n ) \ ,  \ \ \ \
\dot{x} \approx {1\over 2} (x_{n+2} - x_n ) \ .
$$
These expressions can be easily improved to have terms of higher order
but we will not need it.
An elliptic island appears around the point $x = y = 0$ for $\delta < 0$,
i.e. for $K < \epsilon_0$ in correspondence to (\ref{7}) and (\ref{6}). The island
has a scaling
\begin{equation}
y \sim |\delta |^{1/2} x \ , \ \ \ \ (\delta\rightarrow 0) \label{11}
\end{equation}

\begin{figure}
\centering
\includegraphics[width=15 cm]{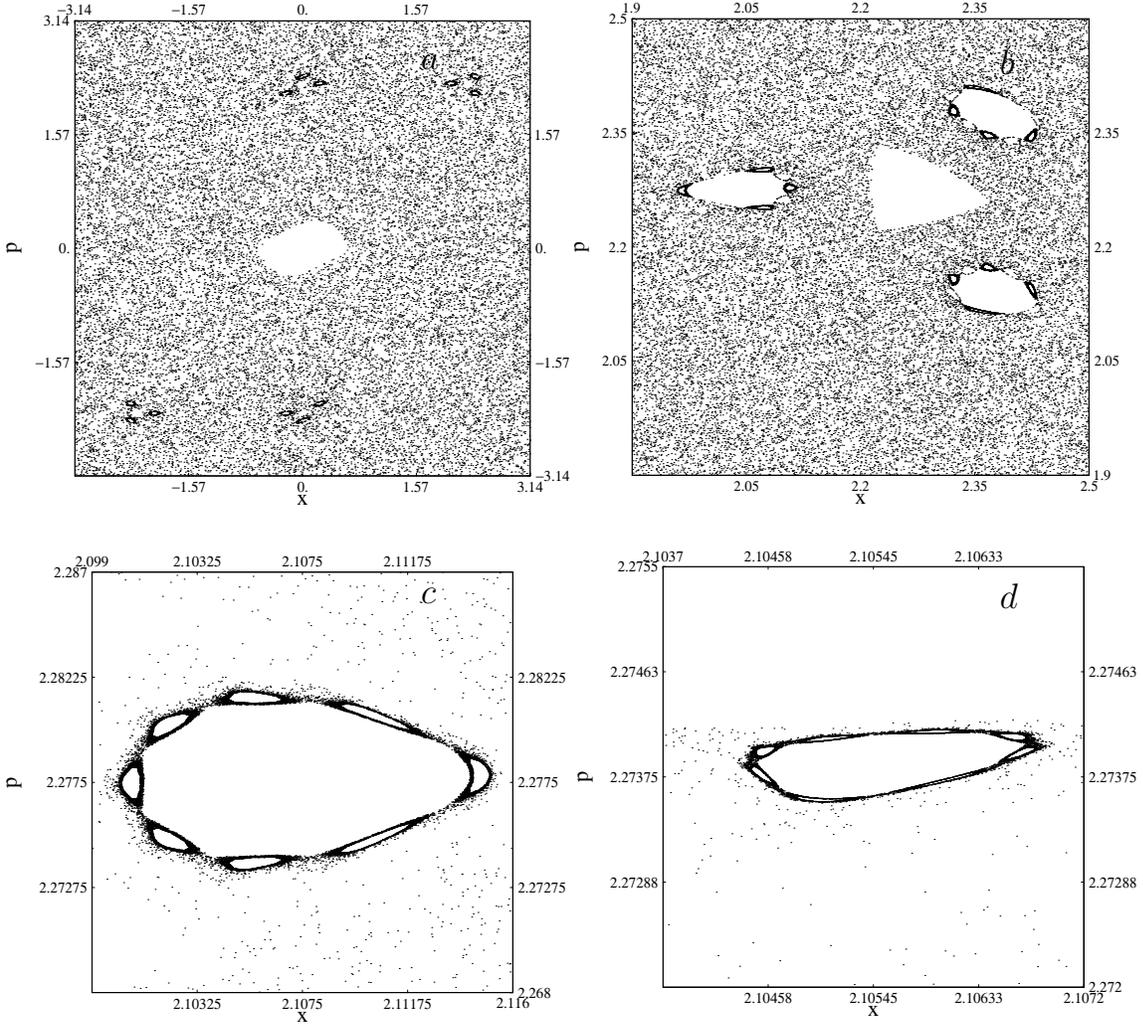}
\caption{\label{fig2}Sticky islands for the same equation as in 
Fig.~\ref{fig1} with $K = K^* = 1$
and $\epsilon_0 = \epsilon_0^* = 1.689538$.
(a) full phase space;
(b) zoom of the group of islands on the top right of (a);
(c) zoom of a next generation island from the left island of (b);
(d) zoom of the lower island in (c). }
\end{figure}
The value $\delta = 0$ is the point of degeneracy and the bifurcation
``opens'' an island in the phase space \cite{Ler}. Increasing of $|\delta |$
leads to a sequence of bifurcations that change the phase space topology.
A particular case, shown in Fig.~\ref{fig2}, is of special interest.
For the
specific values of the pair $(K, \epsilon )$ we have a hierarchical
structure of islands-around-islands with a sequence of islands:
$1 - 4 - 4 \cdot 3 - 4 \cdot 3 \cdot 8 - 4 \cdot 3 \cdot 8 \cdot 8 -
4 \cdot 3 \cdot 8 \cdot 8 \cdot 8 - \ldots$. The values
\begin{equation}
(K^* ,\epsilon^* ): \ K^* = 1 \ , \ \ \ \
\epsilon_0^* = 1.689538\ldots
\label{12}
\end{equation}
are found numerically,
for which the hierarchy of islands has a constant
asymptotic proliferation number 8.
Sequences of islands with bounded values of $m_j : \ m_j \in (m^- ,m^+ )$
is called a hierarchy of islands since the full number of islands at the
hierarchy level $n$ is equal $\bar{m}^n$ with some
$\bar{m} \in (m^- ,m^+ )$ and $n\rightarrow\infty$.

\vspace{.2in}

{\it Property 2.} There exists an infinite number of
hierarchy of islands sequences
$(1, m_1 ,m_2 ,\ldots )$ and the corresponding pairs $(K ,\epsilon_0 )$.
For a fixed
$\epsilon$ (or $K$), it was assumed in \cite{ZEN} (see also
\cite{Z2002}), that the values of $K$ (or $\epsilon_0$) to have an island
hierarchy, are at least as dense as rationals. Many examples of this type
of islands topology can be found in the review \cite{Z2002}.

\noindent
{\bf 4. STICKINESS AND QUASI-TRAPS}

The phenomenon of a long time stay of particles in the vicinity of the
islands border is called stickiness. While it is known that the stickiness
is due to the presence of cantori and some renormalization group approach
can be applied to study particles transport through the cantori \cite{HCM},
a rigorous theory of stickiness does not exist yet and many features of the
stickiness are waiting for our understanding (see more in \cite{Z2002}). A
definition of stickiness can be provided using the notion of dynamical
traps\cite{ZD2002}, or for Hamiltonian systems, quasi-traps.

Let $\Gamma$ be the compact phase space of a dynamical system that is
Hamiltonian and chaotic in a part of $\Gamma$, and let $\Gamma_A \subset
\Gamma$ be an island imbedded in the stochastic sea. Consider a border
of the island, $\partial\Gamma_A$, and an annulus
${\bf A} = \partial\Gamma_A \times {\bf L}$ with a fairly small width $\ell$.
For any initial condition $(x,y) \in {\bf A}$, one can consider a time
$\tau (x,y; {\bf A})$ of a trajectory first exit from $\bf A$.
This time is non-zero and non-infinite, since the dynamics is
area-preserving.

The function $\rho (\tau ; {\bf A})$ is called a distribution function
of the trapping or escape time
from the domain ${\bf A}$  if
\begin{equation}
\Psi (t; {\bf A}) = \int_t^{\infty} d\tau\rho (\tau ;{\bf A}) =
  1 - \int_0^t d\tau\rho (\tau ; {\bf A}) \label{13}
\end{equation}
is the probability to escape from ${\bf A}$ after time $\geq t$. Consider
moments
\begin{equation}
\langle\tau^q \rangle_A = \int_0^{\infty} d\tau \ \tau^q \rho (t; {\bf A})
\ . 
\label{14}
\end{equation}
For $q = 1$ we have $\langle\tau\rangle_A < \infty$ due to the Kac lemma
\cite{Kac}, i.e. the mean trapping time is finite. The domain $\bf A$ is
called quasi-traps if there exists such $q > 1$ that
$\langle\tau^q \rangle_A = \infty$ \cite{ZD2002,Z2002}. Evidently the distribution
function $\rho (t; {\bf A})$ should satisfy the condition
\begin{equation}
\rho (t; {\bf A}) \sim 1/t^{\gamma } \ , \ \  \
\gamma > 1 \ ,
\   \ \ \ (t \rightarrow\infty ) \ ,
\label{15}
\end{equation}
for quasi-traps.

There are many examples of systems with quasi-traps: standard map, Sinai
and Bunimovich billiards (with $\gamma \approx 3$),
rhombic billiard (with, most probably,
$\gamma$ close to 2), and others (see in review \cite{Z2002}). It was assumed
in \cite{Z2002} that realistic Hamiltonian
systems with a hierarchy of islands reveals
the dynamics with quasi-traps. In this paper we
demonstrate by simulations that
similar property exists for the perturbed Anosov diffeomorphism  
(\ref{1}), (\ref{2})
near the point (\ref{12}) where a hierarchy of islands appears.

Visualization of a trajectory of
the map (1) in Fig.~\ref{fig2} shows stickiness as dark strips
around the islands' borders of all presented generations. Although
the stickiness is similar to one, observed in \cite{ZEN}, it is not easy to
follow the same methodology since strong mixing for the Anosov-type
system. In order to provide a quantitative characteristic of the stickiness,
we apply a method of $\epsilon$-separation of trajectories developed in
\cite{AZ,LZ,ZE2005}. We are starting a pair of trajectories in some domain
${\bf B} \subset \Gamma$ fairly far from the islands' area ${\bf A}$. One
trajectory of the pair will be called the reference trajectory, and the
initial distance between trajectories is $d_0 \ll d_{\Gamma}$,
$d_{\Gamma}$ is a diameter of $\Gamma$. The distance at time $t$ is $d(t)$.
Let $t_1$ be a first instant when separation of the trajectories reaches
the value $\epsilon$ that satisfies the condition
\begin{equation}
d_0 \ll \epsilon \ll d_{\Gamma} \ . \label{16}
\end{equation}
After that, at time instant $t_1$ we neglect the trajectory that is not the
reference one and start
another trajectory at a distance $d_0$ from the reference
one. This new pair will be $\epsilon$-separated
first time at $t_2 > t_1$. Such a
procedure continues as long as the reference trajectory runs.

As a result we collect different intervals $\{ \tau_j \} = (\tau_1 ,
\tau_2 ,\ldots )$,
$\tau_j = t_{j+1} - t_j$
of $\epsilon$-separation along the reference trajectory.
The difference between this procedure and the simulation of Lyapunov
exponents is two-fold: it is
in the condition (16), since $\epsilon \ll d_{\Gamma}$, and
the ratio $\epsilon /d_0$ is fixed. Finally, from a large number of the
initial reference trajectories in ${\bf B}$, we obtain a set
$\{ \tau \}_N = \{ \tau_1 ,\ldots ,\tau_N \}$ that provides a frequency
function $\bar{\rho}_N (\tau ; {\bf B} , {\bf A})$
of the relative number of events when $\tau \in (\tau ,\tau + \Delta\tau )$.
If the number of elements
in $\{ \tau \}$ is fairly large, then
$\bar{\rho}_N (\tau ; {\bf B} , {\bf A})$
does not depend on $\bf B$, providing $\Gamma_{\bf B} \cap \Gamma_{\bf A}
= 0$, and it reaches some  limit
\begin{equation}
\bar{\rho} (\tau ; {\bf A}) = \lim_{N\rightarrow\infty}
\bar{\rho} (\tau ; {\bf A} ,{\bf B}) = \lim_{N\rightarrow\infty}
N(\tau ; {\bf A} ,{\bf B}) / N \ ,  \ \ \ \
\int_0^{\infty} d \tau \
\bar{\rho} (\tau ; {\bf A}) = 1 \label{17}
\end{equation}
where
$N(\tau ; {\bf A} ,{\bf B})$
is the full number of $\epsilon$-separations with separation time
$\tau \in (\tau ,\tau + d\tau )$ from the set $\{ \tau \}_N$, and
initial conditions $(x_0 ,y_0 ) \in {\bf B}$.

\begin{figure}
\centering
\includegraphics[width=15 cm]{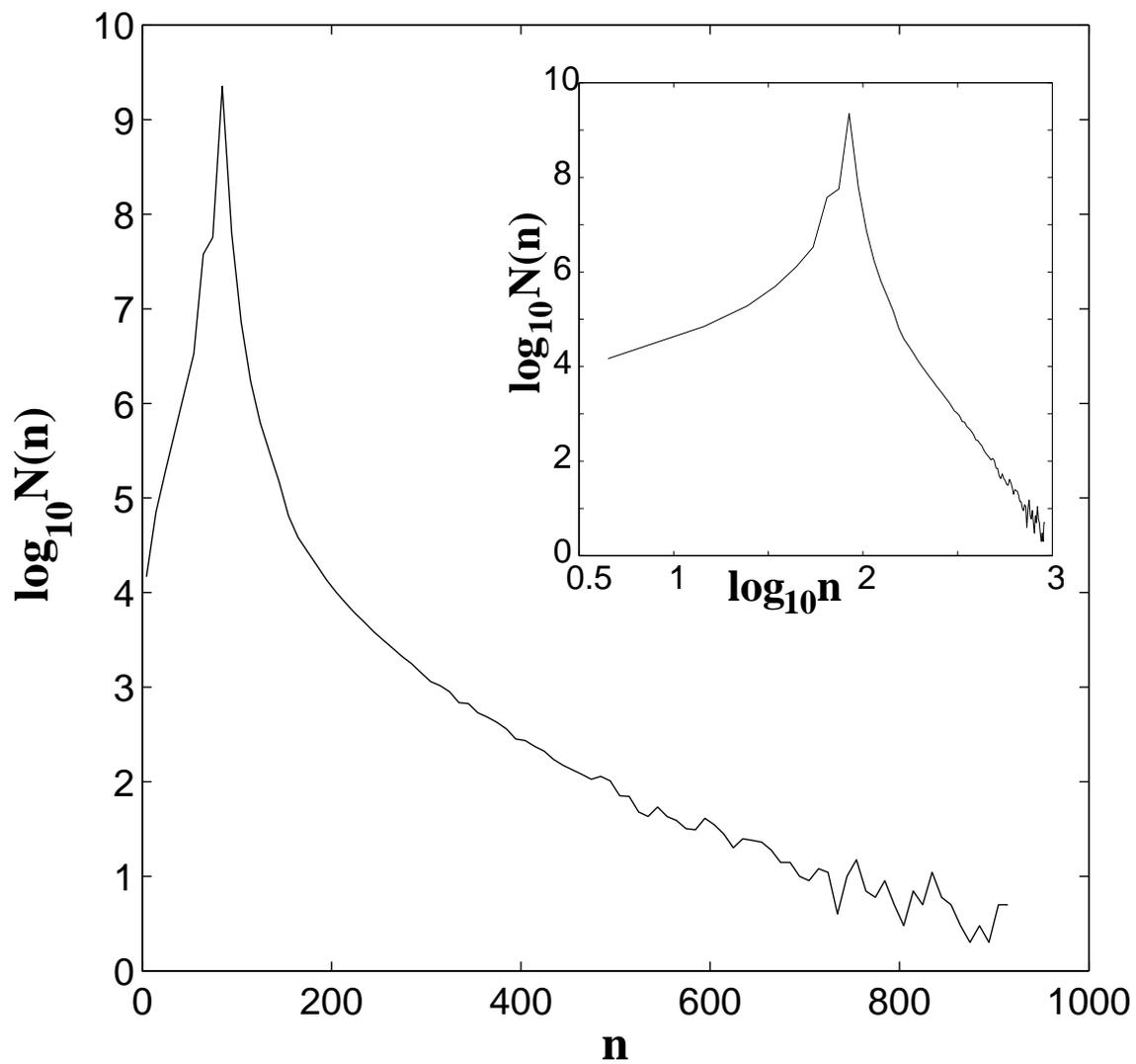}
\caption{\label{fig3} Distribution of the $\epsilon$-separation time intervals for
$K = 0.008$; $\epsilon_0 = 0.008$, obtained from 4048
reference trajectories and $10^7$ iterations each. }
\end{figure}
A typical distribution $N(n)$, where $n$ is discrete time and
${\bf A}$, ${\bf B}$ are omitted, is shown in Fig.~\ref{fig3} 
at the degeneracy
point $K = \epsilon_0$. For the initial distance in pairs we select
$d_0 = 10^{-6}$ and $\epsilon = 10^{-2}$. All 4048 reference trajectories
were run for $10^7$ iterations and full number of the separation events
$N \sim 10^9$. There are 3 different regions in the behavior of $N(n)$.
Before the sharp peak at $n_p \sim 100$ there is 
$N(n) \sim \exp (\bar{\sigma} n )$, where $\bar{\sigma}$ 
is of the order of the mean
Lyapunov exponent (see more below). After the sharp peak, within the
interval $\log_{10} n \in (2, 2.2)$ there is an exponential decay of
$N(n)$, and for $\log_{10} n > 2.3$ the function $N(n)$ is close to the
polynomial decay. The latter statement is provisional due to the lack of
statistics and small time interval. Because of inhomogeneity of phase space,
the finite time Lyapunov exponent $\sigma_t$ has a distribution function,
and the first two domains with the peak fit the corresponding theory of the
probabilistic analysis of the distribution of $\sigma_t$ (see for example
in\cite{Ott}).
The third domain shows a new feature that can be interpreted
as influence of the bifurcation point that leads to the appearance of the
stickiness near $(x,y) = (0,0)$.

\begin{figure}
\centering
\includegraphics[width=15 cm]{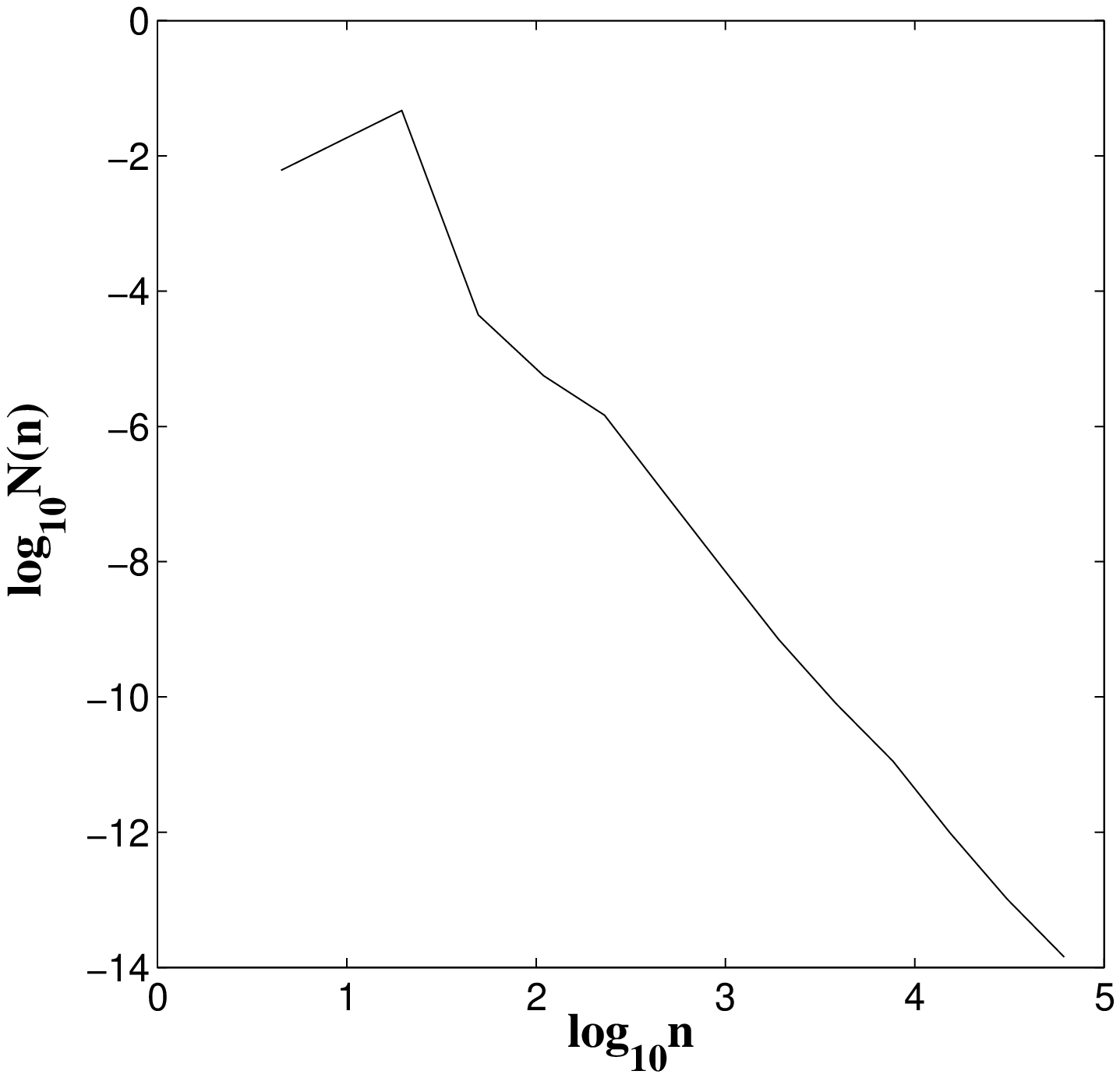}
\caption{\label{fig4} The same as in Fig.~\ref{fig3} but for special values $K^* = 1$,
$\epsilon_0^* = 1.689538$.}
\end{figure}

Another situation occurs for the values of parameters $(K^* ,\epsilon^* )$,
given in (\ref{12}) when the stickiness to the hierarchical islands is strong.
For this specific case, let us denote
\begin{equation}
P(n) = \lim_{N\rightarrow \infty} N(\tau ,{\bf A}, {\bf B}) /N \label{18}
\end{equation}
(compare to  (\ref{17})). A corresponding result is shown in Fig.~\ref{fig4}. 
Within four
decades of $n$ there is a clear power-wise behavior of $P(n)$:
\begin{equation}
P(n) \sim {\rm const.} \ n^{-\gamma } \ , \ \ \ \ \
(n\rightarrow\infty ) \label{19}
\end{equation}
with $\gamma \approx 3.3$. To analyze this result, we need to

\vspace{.2in}

{\it Conjecture} \cite{ZE2005}:
Assume that the phase space $\Gamma$ of compact dynamics can be split in
two parts: ${\bf A}$ that corresponds to the only sticky (singular) zone
and $\Gamma \backslash {\bf A}$, and that
${\bf A} \cap \Gamma \backslash {\bf A} = 0$.
Then for fairly small ${\bf B} \subset \Gamma \backslash {\bf A}$
the following
equality is valid
\begin{equation}
\gamma_{\rm rec} = \gamma \ , \ \ \ \ \
(t\rightarrow\infty ) \label{20}
\end{equation}
where $\gamma_{\rm rec}$ is the recurrence exponent of the distribution of
Poincar\'{e} recurrences
to ${\bf B}$
$$
P_{\rm rec} (n) = {\rm const.} \ n^{-\gamma_{\rm rec} }
$$
\begin{equation}
\sum_{n=1}^{\infty} P_{\rm rec} (n) = 1 \ . \ \ \ \
(n\rightarrow\infty ) \label{21}
\end{equation}
In other words, distribution of $\epsilon$-dispersion of trajectories for
small $\epsilon$ is defined asymptotically by the sticky zone.

\begin{figure}
\centering
\includegraphics[width=15 cm]{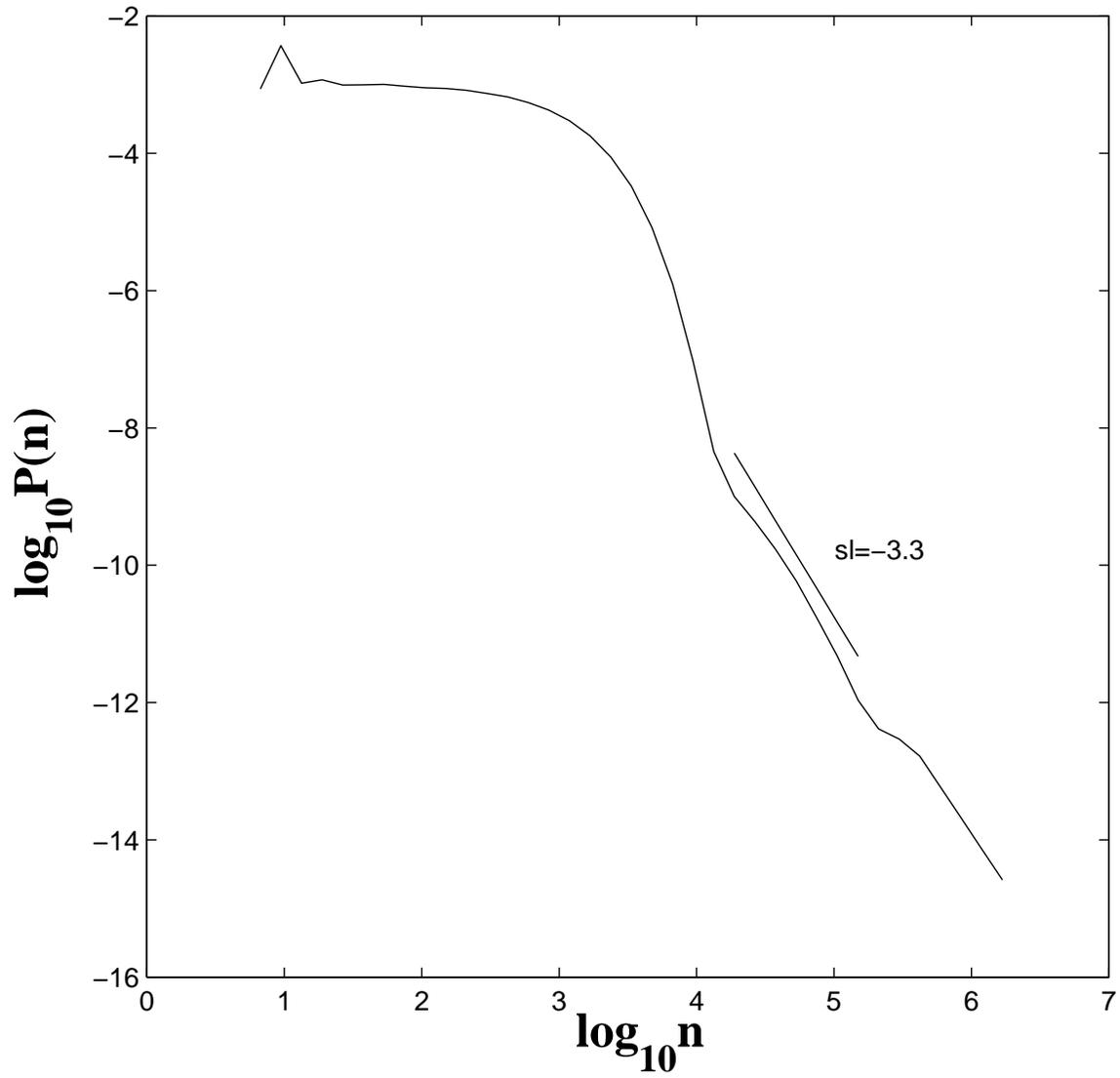}
\caption{\label{fig5} Distribution of
Poincar\'{e} recurrence
time for special values $K^* = 1$, $\epsilon^*_0 = 1.689538$}
\end{figure}
The value $\gamma_{\rm rec} = 3.3$ may be of a universal character since
similar value was obtained for the standard map in \cite{ZEN}. It is
worthwhile to check the statements (\ref{20}) and  (\ref{21}) straightforward by
calculating the distribution of
Poincar\'{e} recurrences.
The corresponding data are given in Fig.~\ref{fig5}. For the time interval
$n < 10^4$
\begin{equation}
P_{\rm rec} (n) \sim \exp (-n/\bar{n} ) \label{22}
\end{equation}
but for $n > 10^4$ there is a crossover to the power law  (\ref{21}) with the
same exponent $\gamma_{\rm rec} = 3.3$.
\newpage
\noindent
{\bf 5. CONCLUSIONS}

Our numerical results demonstrate that the perturbed Anosov map shows the
features of the typical realistic Hamiltonian systems of low dimension:
appearance of islands as a result of a bifurcation; sequence of
topological changes in the phase space when we change parameters of the
system; islands-around-islands hierarchical set; stickiness of the
islands' boundaries. The most important new features of the perturbed Anosov
map is the transform of the
Poincar\'{e} recurrences
distribution from the exponential decay to the power-wise decay as
$t \rightarrow\infty$. Such a change is important for numerous applications
being responsible for different anomalous properties of the observable
macroscopic characteristics \cite{Z2002}.

Both, Property 1 and Property 2 formulated in Sec.~3, include as their main 
part the existence of different infinite hierarchies of islands. These 
statements are conjectures based on high accuracy simulations. The area of 
stickiness of orbits can be characterized by a low rate of dispersion of 
trajectories. That means that the long time, necessary for  
trajectories separation, can be effectively lower if the accuracy is 
insufficient. A possibility to observe a crossover from exponential dispersion
of trajectories to the power one just confirms the high level of accuracy of 
simulations. To be more specific, let us mention that the critical value of 
perturbation $\epsilon_0^*$ is defined up to six decimal digits in (\ref{12}).
The crossover from the exponential decay of $P(n)$ to the power one occurs at 
$ \log_{10}n \approx 4$ (see  Fig.~\ref{fig5}) and the power behavior of 
$P(n)$ is up to $ \log_{10}n \geq 6$. These results are consistent with level 
of accuracy of $\epsilon_0^*$. 
\vspace{.3in}

\noindent
{\bf Acknowledgments}

We are very grateful to L.M. Lerman for numerous discussions and a possibility
to read his notes prior to publication. This work was supported by the
Office of Naval Research Grant No. N00014-02-1-0056, and the U.S.
Department of Energy Grant No. DE-FG02-92ER54184. Simulations were supported 
by NERSC.

\end{document}